# Bringing Reference Groups Back:

# Agent-based Modeling of the Spiral of Silence

Cheng-Jun Wang

Web Mining Lab, Dept. of Media and Communication, City University of Hong Kong

Address Correspondence to:

Cheng-Jun Wang

Run Run Shaw Creative Media Centre, City university of Hong Kong.

Kowloon, Hong Kong.

Email: wangchj04@gmail.com

**Abstract**

The purpose of this study is threefold: first, to bring reference groups back into the framework of spiral of silence (SOS) by proposing an extended framework of dual opinion climate; second, to investigate the boundary conditions of SOS; third, to identify the characteristics of SOS in terms of spatial variation and temporal evolution. Modeling SOS with agent-based models, the findings suggest (1) there is no guarantee of SOS with reference groups being brought back; (2) Stable existence of SOS is contingent upon the comparative strength of mass media over reference groups; (3) SOS is size-dependent upon reference groups and the population; (4) the growth rate of SOS decreases over time. Thus, this research presents an extension of the SOS theory.

*Key Words*: spiral of silence; willingness to express; agent-based model; reference group; size-dependent

**Bringing Reference Groups Back:**

**Agent-based Modeling of the Spiral of Silence**

**Introduction**

Spiral of silence (SOS) theory contributes the research of public opinion by uncovering the interplay between individuals and mass media. However, there are many debates on the fundamental requisites of the theory, and the inconsistent findings of empirical studies intensify these criticisms (Donsbach & Traugott, 2007; C. J. Glynn, Hayes, & Shanahan, 1997; Scheufle & Moy, 2000).

One critique of SOS studies is the absence of reference groups (Donsbach & Stevenson, 1984; C.J. Glynn & McLeod, 1984; C. J. Glynn & Park, 1997; Kennamer, 1990; Salmon & Kline, 1983; Salwen, Lin, & Matera, 1994). Noelle-Neumann (1984) outlines two source of influential factors for individual perceptions about the majority opinion: "firsthand observation of reality and observation of reality through the eye of the media" (p. 159), based on which, the idea of dual climate of opinion has been touched by Noelle-Neumann (1984) in the ground-breaking book of *The Spiral of Silence*. However, Neumann (1984) views it as "an unusual weather situation or a distant vista" which seldom happens (p. 168). Thus, in later research, this conceptualization of dual climate of opinion was not well incorporated into the formal model of SOS.

The theory of SOS may overestimate media influence, since each individual is viewed as an atom, while the reality is individuals have their reference groups. Reference group was firstly used by Hyman (1942) to denote the group to which individuals relate their attitudes. Reference groups are presumed to supply references to individuals for judgment (Shibutani, 1955), and deliver punishments or rewards to group members (Kelley, 1952). With more and more studies confirming the significant influence of reference groups on SOS (Bowen & Blackmon, 2003; C. J. Glynn & Park, 1997; Moy, Domke, & Stamm, 2001; Neuwirth & Frederick, 2004; Oshagan, 1996; Salwen, et al., 1994), the underestimated opinion climate constructed by reference groups deserves being reconsidered.

The general research questions of this paper concern the boundary conditions of SOS. Using agent-based models (ABMs) to integrate factors of individual level (e.g., willingness to express), group level (e.g., reference groups), and societal level (e.g., mass media), this study aims to bring reference groups back into the SOS theory, propose an extended framework of dual opinion climate, and explore the features of SOS in terms of spatial variation and temporal evolution.

## The Spiral of Silence Theory

The SOS theory starts from Noelle-Neumann's puzzlement about German elections in 1965 and 1972. While voting intentions kept almost constant, Noelle-Neumann (1974, 1984, 1993) finds that the expectations of who was going to win the election change dramatically. To address this question, Noelle-Neumann (1974) formulates the hypothesis of SOS. To be specific, the society isolates deviant individuals, and individuals continuously experience the fear of isolation. To avoid being punished by the society for holding minority opinion, individuals actively monitor the opinion climate, estimate the distribution of public opinion using the quasi-statistical sense, and form their perceptions of the majority opinion. If they feel their opinions are losing grounds, e.g., if they think their opinions are different from the dominating ideas of mass media, their willingness to express will decrease, and they will fall silent, which will reinforce their perceptions about the majority opinion, and gives rise to the presence of SOS.

There are mainly two lines of research about SOS in terms of dual opinion climate. One focuses on testing and verifying the underlying assumptions of SOS within the classic framework to address the theoretical issues, including: perceived media opinion (Kim, Han, Shanahan, & Berdayes, 2004), fear of isolation (Lin & Pfau, 2007; Neuwirth, Frederick, & Mayo, 2007; Petri & Pinter, 2002; D. A. Scheufele, Shanahan, & Lee, 2001), pluralistic ignorance (Taylor, 1982), hardcore (C.J. Glynn & McLeod, 1984), culture differences (Huang, 2005; Lee, Detenber, Willnat, Aday, & Graf, 2004; McDonald, Glynn, Kim, & Ostman, 2001). Another substantial body of literature argues that, despite of societal-level opinion climate, local-level climate structured by reference groups is more important (Donsbach & Stevenson, 1984; C.J. Glynn & McLeod, 1984; C. J. Glynn & Park, 1997; Oshagan, 1996; Salmon & Kline, 1983; Salmon & Oshagan, 1990; Salwen, et al., 1994).

In the research of SOS, reference groups have been argued to play a crucial role in opinion formation and maintenance. First, reference groups shape the local opinion climate which individuals directly experience in their daily life. E.g., Salmon (1983) argues that individual perception of "social reality" is structured by reference groups. Second, reference groups act as a proxy to link different factors on multi-levels. The group-level communications link the individual-level perceptions and the societal-level impacts (e.g., media effect and culture influence). This claim has been warranted by diverse perspectives, including opinion leaders (Katz, 2002; Lazarsfeld, Berelson, & Gaudet, 1968), structural model of mass communication (Tichenor, Donohue, & Olien, 1973), threshold models (Granovetter & Soong, 1983, 1986, 1988; Krassa, 1988), social impact theory (Nowak, Szamrej, & Latane, 1990), theories of planned behavior (Neuwirth & Frederick, 2004).Third, reference groups compete with mass media to influence individuals' perceptions about dominant opinion. According to Gallup et al. (1940) and Salmon et al. (1983), reference groups plays more important role than national opinion in forming individual's perception of the majority opinion. Krassa (1988) states that a main influence on willingness to express is the perceived distribution of opinion among the reference groups. Kennamer (1990) asserts that SOS will occur if people don't have supports from peer groups and feel hostile opposition to their opinions.

      Following the arguments above, more and more studies have confirmed the significance of reference groups. Salwen et al. (1994) find that people would like to speak out when their opinions are congruent with local climate rather national climate. Oshagan (1996) finds that when reference group and social majority opinion are made equally salient, the influence of reference group is more important. Glynn et al. (1997) find the indirect influences of reference group on SOS. Moy et al. (2001) further demonstrate that perceived opinion with family and friends—rather than society at large—predicted willingness to speak out. The study of Bowen et al. (2003) demonstrates that individuals in organization are more likely to speak out when they find supports from their workgroups. Neuwirth et al. (2004) compare peer influences in terms of theories of planned behavior with perceptions of majority attitude derived from SOS theory, and finds that peer opinions have more influence.

### Extended Model of Spiral of Silence

      To summarize, it becomes more and more significant to bring reference groups back.

In terms of dual climate of opinion, this present article attempts to extend the classic theory of SOS by including reference groups. I propose an extended model (see Figure 1), and claim that dual opinion climate shapes individuals' perceptions about the majority opinions.

As it has been asserted by Scheufele (2008), dual climate of opinion links micro-, meso-, and macro-levels of analysis. Accordingly, there are at least three levels of SOS, including individual-level, group-level, and societal-level. First, as a micro-theory, SOS examines people's willingness to express, fear of isolation, quasi-statistical sense, and the demographic attributes on individual level. Second, as a meso-theory, SOS emphasizes the influences of reference groups. Following this line of though, the size, opinion climate, and resources of reference groups are expected to have impact on individuals' perceptions about the majority opinion. As the proxy of individual-level factors and societal-level impacts, reference group turns to be more important in terms of the transmission of both information and influences. Third, as a macro-theory, SOS emphasizes the influence of mass media on the societal-level. The direct influence of mass media as a societal-level factor on individuals' perception of the prevailing opinions is warranted by the ubiquity effect, the consonant effect, and the accumulation effect of media influence (Donsbach & Traugott, 2007; Gonzalez, 1988). The opinion expressed by mass media comprises the national opinion climate. In this sense, mass media is one kind of social control.

_______________

Insert Figure 1 here

_______________

Another newingredient of this extended model is opinion threshold. The widely cited articles dealing with racial dynamics by Schelling (1971, 1972) introduces threshold into later studies. He shows that individuals have their tipping points for a mixture of residents' colors. If the proportion of residents of different colors living in the neighborhood exceeds individuals' thresholds, they will move out of the area to avoid being in the minority. Following Schelling's steps, Granovetter et al. (1983, 1986, 1988) formally propose threshold models which assume that individual behavior depends on the number of people who have already engaged in that behavior. Threshold has been used to study interpersonal effect of collective behavior, e.g., residential segregation (Granovetter & Soong, 1988; Schelling, 1971), consumer demand (Granovetter & Soong, 1986), and diffusion of innovations (Valente, 1996). In the study of SOS, opinion threshold which is defined as the levels of public support that an individual requires to speak out (C. J. Glynn & Park, 1997; Krassa, 1988). Unless individuals' willingness to express exceeds a threshold, they will not speak out or fall silent.

   Put together, as a dynamic process of SOS, individuals learn about the opinion climate by observing directly from the reference groups, and the climate portrayed by the media. As it has been demonstrated in Figure 1, the dashed line denotes individuals. They are oppressed by the threat of isolation from the society. Individuals monitor mass media to relieve the fear of isolation. However, they are not alone, since they have reference groups whereby they obtain information and supports or receive punishments. Through the quasi-statistical sense, individuals perceive and evaluate the dual opinion climate (e.g., the distribution of opinions), and according to their own opinion thresholds, they choose to speak out or not.

   The central concern of this study is to examine how individuals respond to the dual opinion climate which gives rise to the emergence of SOS. Two competing factors, i.e., mass media and reference groups, will be compared.

First, according to the theory of SOS in terms of societal threat, media opinion indicates the majority opinion of the public. Individuals' reactions to the mass media demonstrate how the social structure may influence the individual behaviors. Thus, media influence would definitely reinforce the spiral process.

Second, the influence of reference groups is dual. Reference groups will punish the individuals if their opinions are different from reference groups, or else, it will reinforce individual opinions (Kelley, 1952). For those reference groups which hold the opinion of the mass media, they will reinforce the influence of the mass media. On the contrary, the people holding the opinion of against that of the mass media will turn to the reference groups for supports. Thus, reference groups may decrease or even reverse the influences of mass media. It's straightforward to integrate the thoughts above, and propose the following questions:

**RQ1**: How does reference group influence individuals' willingness to express?

**RQ2**: What's the boundary condition of a stable existence of SOS after reference groups being brought back?

Social interactions on both local scope and societal-level are constrained by social structures, especially the size of population, and the size of reference groups. According to Nolle-Neumann (1994), fear of isolation increases in proportion to the size of the publics. Similar arguments hold for reference groups. It leads us to postulate that SOS is size-dependent on population and reference groups. Therefore, it's reasonable to identify how the size of population and the size of reference groups influence SOS. To be specific:

**RQ3**: How does the size of reference groups influence the emergence of SOS?

**RQ4**: How does the size of population influence the emergence of SOS?

Investigating social process with time dimension will shed more light on our understanding about opinion dynamics (Allport, 1937). It's significant to gauge the dynamic process of SOS, especially how the number of people falling silent evolves over time. Thus, I formulate the last question for this article.

**RQ5**: How does the number of people falling silent change over time?

## Method

Heterogeneous individuals interplay with mass media and reference groups at the local scope give rise to the SOS as an emergence of macroscopic regularity. Just as Scheufele (2008, p. 182) has stated, "most importantly, however, future studies will have to examine the interactions between these aggregate-level differences and the individual-level predictors of outspokenness". Unfortunately, most statistical models adopted in the past research fail in capturing the dynamic interplays between aggregate-level factors and individual-level predictors (Donsbach & Traugott, 2007).While, by linking the micro-level, meso-level, and macro-level, agent-based models deal with these challenges well.

**Agent-based Models (ABMs)**

As one approach of computational social science, ABMs enable modelers to create, analyze, and experiment with models composed of agents who interact within an environment (Gilbert, 2008). ABMs began with Von Neumann's work on self-reproducing automata (Neumann & Burks, 1966), then it is widely used in physics, mathematics, biology, etc. Its applications in social science are promoted by Conway's model of "game of life" (Gardner, 1970), Schelling's model of neighborhood segregation (Schelling, 1971), and Axelrod's model of tit for tat (Axelrod & Hamilton, 1981). ABMs also give rise to abundant research focusing on opinion dynamics (Suo & Chen, 2008; Weisbuch, Deffuant, & Amblard, 2005; Weisbuch, Deffuant, Amblard, & Nadal, 2002) and diffusion of innovations (Bullnheimer, Dawid, & Zeller, 1998; Rosenkopf & Abrahamson, 1999; Strang & Macy, 2001).

ABMs are designed to explore the minimal conditions or assumptions required by specific social phenomenon which emerges at a higher level of organization (Macy & Willer, 2002), therefore, it is helpful for us to identify integrated mechanisms by tracing the micro-macro links, and capture the domino effect, emergence, and criticality in social process (Squazzoni, 2008).

In ABMs, agents are distributed in the patches of the network. Both agents and patches have specific properties. Agents interact with each other and the environment following simple rules. They are interdependent with each other, adaptive and backward-looking (Macy & Willer, 2002). With the behavior of different agents being governed by a set of rules, certain properties may emerge on the macro-level (Nowak, et al., 1990). For more information about ABMs, readers can refer to Gibert's book *Agent-based Models* (Gilbert, 2008).

**Measure**

Although ABMs employ simulations, they do not aim to provide an accurate representation for a particular empirical reality. Instead, the goal of agent-based modeling is to enrich our understanding of fundamental processes that may appear in a variety of applications (Axelrod & Hamilton, 1981). There are many factors that will affect SOS, and this paper just focuses on three constructs according to three levels of analysis: the willingness to express, the influence of reference groups, and the influence of mass media. I denote media influence with α, and group influence with β. In addition, I specify the measurement for intensity of group involvement and frequency of media use.

**The willingness to express & opinion threshold**. Each agent begins with the personal preference of willingness to express which measures people's choice of speaking out or not. It's a continuous variable deprived from a uniform normal distribution. The mean of the normal distribution is 0, and the standard deviation of the distribution is 1. I define the agents holding negative willingness to express as silent people, and agents with positive values as those who tend to speak out, even though their opinions are different from the dominant opinion (e.g., media opinion).  Thus, the mean value of willingness to express, in this case is 0, serves as opinion threshold. Driven by dual opinion climate, willingness to express changes over time. People who find opinion climate is hostile to their own opinions *tend* to fall silent to avoid being isolated (C. J. Glynn, et al., 1997; Noelle-Neumann, 1974; Scheufle & Moy, 2000).

**Intensity of group involvement**. Each agent locates in its social network, and gets supports from the reference groups. According to Noelle-Neumann (1974), agent has a "quasi-statistical sense" which can accurately evaluate the opinion climate. Each agent has a radius of observation which measures the size of the reference groups. Different from the assumption which assumes that individuals can monitor all the people in the society, in agent-based models, the agents are assumed to be only bounded rational. Accordingly, agents can only experience the dual opinion climate in the local scope. To compute the intensity of group involvement for agent i, the willingness to express of each agent in i's reference groups (i.e., in i's radius of vision) are summed up.

**Frequency of media use**. Since media influence is ubiquitous, it is not need to set specific agents as mass media. Thus, I model frequency of media use by the property of the patches. Each patch has a property named media exposure with the value randomly taken from a list (0, 1, 2, 3, 4, 5). 0 indicates there is no media exposure on the patch, from 1 to 5, the bigger the number is, the bigger the media exposure on the patch is.

One important influence comes from "hard cores" and avant-gardes", both of which resist the dominant opinion climate (Donsbach & Traugott, 2007; Noelle-Neumann, 1984). To include them in, I set some places without media use (i.e., the frequency of the media use equals 0). Note that, when hard cores and avant-gardes have the same opinion with reference groups, reference groups will provide a protective environment for them. But when the influence of the reference groups are overridden by the mass media, i.e., the opinion of reference groups is congruent with the mass media, and different with hard cores and avant-gardes, reference groups will compel hard cores and avant-gardes to fall silent.

**How Does Agent-based Model Work?**

I use NetLogo 4.13 to model SOS. Netlogo is a programmable modeling environment authored by Uri Wilensky. It is suitable for modeling complex systems developing over time to explore the connection between the micro-level behavior of individuals and the macro-level patterns (Sklar, 2007; Wilensky & Rand, 2009).

Agents act in respect with behavioral rules over time t, the number of agents is N, the willingness to express is expressed by W, the nth agent's willingness to express at time t is expressed as $W_{n,t}$, the nth agent stays at the nth patch, the frequency of media use at the nth patch at time t is $M_{n,t}$, the nth agent's opinion climate of reference groups at time t is expressed as $R_{n,t}$, and the coefficient of $M_{n,t}$ is expressed as α, the coefficient of $R_{n,t}$ is expressed as β. Thus, the nth agent's willingness to express at time t is demonstrated in formula (1):

$$W_{n,t} = W_{n,(t-1)} + \alpha M_{n,(t-1)} + \beta R_{n,(t-1)} \quad (1)$$

At the initial stage, let 1000 agents randomly distribute in the environment, and determines the boundary of each agent's reference groups by fix its radius of vision be 3. Hereafter, we can alter the radius of vision to control the size of reference groups. To reproduce the neck-to-neck competition between two sides, and to simplify the simulation, I set half of the agents agree with media opinion (or fall silent), and half of them not, in the initial stage of simulation. Since agent-based model is also one kind of experiment, it needs to control the influence of each factor. Given specific conditions, it's easy to test the group influence and media impact, respectively.

## Results

**The Influence of Mass Media**

This research starts from identifying the impact of mass media. To illustrate the media influence, I specify the null model as following: the population size =1000, the vision = 3, β = 0, α = 0.02, and I run the experiments for 100 times. The findings reveal that, driven merely by mass media over time, there is a clear pattern of linear growth for the number of silent agents ($\beta = 0.997$, $R^2 = 0.993$, $p < .001$). Media effect is very obvious and stable: it just makes more and more people fall silent, which supports the idea that public opinion shaped by mass media can be viewed as a kind of social control (Noelle-Neumann, 1984, 1993; Scheufle & Moy, 2000).

**The Influence of Reference Groups**

RQ1 concerns the influence of reference groups. Given the population size = 1000, vision = 3, β = 0.02, α = 0, I run the experiment for 100 times, and find the result is mixed and unstable. Without societal level impact, both SOS (48%) and "spiral of speaking-out" emerge (52%).

Further, the findings indicate that the agents who are close with each other will converge (see Figure 2). After several steps of contagion, the silent people and the talking people are separated into two clusters. Two sides combat with each other on their boundaries, and the strong side dominates the opinion by surrounding and isolating the weak side. Without the influence of mass media, the convergence of competing opinions is based on agents' geographical distribution. It's crucial to put the right agent in the right place, which is like the game of chess. Thus, the evolution of opinion dynamics driven merely by reference groups is path-dependent upon the structure of social networks.

_______________

Insert Figure 2 here

_______________

**The Boundary Conditions of Spiral of Silence**

RQ2 is about the major concern of this paper, i.e., the boundary conditions of SOS.

Reference groups compete with mass media to influence opinion dynamics. I denote the media influence with α, and group influence with β. Thus, the ratio of α/β which measures the relative strength of mass media relative to reference groups, supplies a good way to identify the boundary conditions for SOS. Following this logic, I define the reference groups as strong reference groups when α/β < 0.1, and strong media when α/β > 10. To address the first question, I set radius = 3, N = 1000, α/β equals 0.1 and 10, and run each model for 100 times to compare the differences.

For the situation of strong media, I establish the model by setting α/β=10; radius=3; N = 1000. The global regularity of SOS stably emerges from individual behaviors. While, for the situation of strong reference groups (e.g., α/β = 0.1), the results are unstable. Similar to the social setting with only group influence, I find that both SOS (62%) and the spiral of speaking-out (38%) emerge.

Further, given a social setting of strong media, I want to specify how reference groups influence the process of SOS. To do so, I conduct two experiments by setting α = 0.002, β = 0.0001 for control group, and α = 0.002, β = 0.0005 for experimental group. The results demonstrate that control group takes 52.80 ticks to converge, and experimental group takes 43.55 ticks to converge. Thus, given a stronger mass media relative to the strength of reference groups, the population with only "strong" reference groups falls silent even faster than that with weak reference groups ($t(37.4) = 10.7, p < .001$), which implies that when reference groups are overridden by mass media, reference groups will reinforce the media impact.

Put together, the boundary condition of SOS is the comparative strength between mass media and reference groups. Given the boundary condition of strong reference groups (compared with media influence), there is no guarantee for the stable existence of SOS. While, if media influence is much larger than group influence, reference groups will reinforce media influence for creating SOS.

**The Size of Reference Groups**

RQ3 is about the size of reference groups. To answer RQ3, I set $\alpha = 0.002$, radius = 3, $N = 1000$, $\beta = 0.0001$, and adjust the vision of the agents as 2, 4, 6. When the vision is 2, the agents only take account of the opinion of the agents who are in the radius of 2 patches. The larger the vision is, the larger the size of reference groups is, and the more information agents can get from the neighborhood.

The average convergence time for three groups is 162.79, 126.19, and 98.02, respectively, i.e., the larger size of reference groups is, the faster the population reach consensus. A one-way ANOVA was used to test for differences among three sizes of reference groups, which reveals that the convergence time differes significantly across the three sizes, $F(1, 298) = 2524.5, p < 0.001$.

**The Size of Population**

RQ4 concerns the influence of population size. I set the population of the agents as 1000, 1500, 2000, and run the experiments 100 times for each situation. The results indicate that average convergence time for three groups is 161, 156, and 98, respectively. A one-way ANOVA was used to test for differences among three sizes of population, which reveals that convergence time differed significantly across the three sizes, $F(1, 298) = 774.67, p < 0.001$. Thus, the larger the population size is, the sooner people fall silent.

**The Evolution of Opinion Dynamics over Time**

To answer RQ5, I explore how the number of people falling silent each time evolves with time, with the simulation data. The result indicates that the number of people falling silent each time is negatively correlated with time ($r(7970) = 0.746$, $p < 0.001$). Thus, the cumulative number of people falling silent grows fast at first, but the number of people falling silent each time gets smaller over time.

## Discussion and Conclusion

This study views SOS as a global regularity emerging from individual interactions with dual opinion climate over time. The primary findings of this study focus on the function of reference groups. Individual interactions with reference groups at the local scope may reinforce media effect on SOS, but it can also countervail or even reverse SOS. The direction of the influence is contingent upon the comparative strength of mass media over reference groups, which is the boundary condition of SOS.

This finding corresponds to the primary function of reference groups: it tends to punish the individuals whose opinions are different from reference groups, or else, it tends to reinforce individuals' behavior if their opinions are similar to reference groups (Kelley, 1952). Accordingly, SOS and pluralistic ignorance hold only when individuals lose supports of reference groups, or the influence of reference groups have been overshadowed by mass media. Or else, there is no guarantee of SOS, especially when the influence of mass media is smaller than that of reference groups ($\alpha/\beta < 1$).

It is necessary to note that Noelle-Neumann (1984) had been aware of the boundary conditions of SOS. Actually, Noelle-Neumann (1984) acknowledges that the dual climate of opinion is a fascinating phenomenon which may act as a "struggle against the spiral of silence" (p. 167), although she asserts that dual opinion climate can only arise under very special circumstances, especially "when the climate of opinion among the people and that dominant among media journalists diverge" (p. 168).

Within the boundary conditions, this study probes the influence of reference groups. Note that, reference groups are overridden by mass media in this situation, therefore reference groups tends to hold the same opinion of mass media, and compel agents to fall silent. As a proxy, reference groups will spread the media influence further. The stronger reference groups are, the sooner SOS emerges; the larger reference groups and the population are, the sooner SOS presents.

The findings shed light to the debate about the role of reference groups in SOS. For example, Krassa (1988) asserts that the more socially integrated the community is (e.g. the stronger reference groups are), the less we need worry about the SOS (e.g., the tyranny of the majority), since in a densely connected society or community, people are more sensitive to the actions and opinions of others, therefore, it's easier to mobilize or demobilize the community against the tyranny (Krassa, 1988). According to the findings of this research, Krassa's assertion is be right when group opinions are against media opinion, and group influence is not smaller than media influence. While, if reference groups have been overridden by mass media, the more socially integrated, the easier to make the population fall silent.

Similarly, Salmon et al. (1990) postulated that large community is more diverse in opinions. Thus, individuals from larger communities feel less certain about the majority opinions, which makes it harder for the mass media to make those people reach consensus (Salmon & Oshagan, 1990). This assertion is right when reference groups are at least not weaker than mass media. While, if the opinion of reference groups has been overridden by the mass media, the larger the population is, the sooner SOS emerge, which is congruent with the assertion that the fear of isolation increases with the size of the public (Noelle-Neumann, 1993). Thus, it's necessary to be aware of the boundary conditions of SOS.

In all, this paper contributes to our knowledge about the interplay among individuals, reference groups, and mass media. First, bringing reference groups back to SOS, and extend the model of SOS in terms of dual opinion climate and opinion threshold;  Second, agent-based models supplies one alternative choice to analytically investigate the boundary conditions of SOS, which serves as one way to validate the internal validity of SOS theory. While the reference groups are brought back, there is no guarantee of SOS: (1) when the opinion of the reference groups is different with media opinion, reference groups may countervail, stop, or even reverse the trend of SOS; (2) when the opinion of reference groups is similar to or overshadowed by media opinion, it will reinforce SOS. Third, this study highlights the dynamic features of SOS: (1) SOS is size-dependent upon reference groups and the population; (2) the growth rate of SOS decreases over time.

This paper extends our understandings of the functions of reference groups. Especially, the role of proxy reference groups play, and the reinforcement effect of reference groups in a social setting of strong media and weak reference groups. The logic has the potential to explain the other patterns of public opinion. For instance, Brosius and Kepplinger (1990) studied the German television and news, and found that agenda setting is most pronounced when individuals have no direct contact with an issue and thus are dependent on the media for information. Following this logic, Weimann and Brosium (1994) proposed the agenda-setting is also a process of two-step flow. Forth, the growth rate of the number of people falling silent per time, which is driven by the dual climate of opinion, diminishes over time. Another case is the cultural impact on SOS which is thought to be the most fruitful area in current research of SOS (Donsbach & Traugott, 2007; Scheufle & Moy, 2000). This line of research underscores the function of reference groups, since the emphasis of collectivism and individualism is closely relevant to the features of reference groups (Chen & West, 2008).

This research is not without limitations. For example, I use the lattice network as the environment to test SOS, while the real social network maybe small-world network (Watts & Strogatz, 1998) or scale-free network (Barabasi & Crandall, 2003). Thus, it's necessary to test and verify the conclusions in scale-free network and small world network in the future. However, compared with chasing mixed findings of SOS in concrete social settings, it's encouraging to have a generalized formal model to capture the underlying mechanisms of SOS.

BRINGING REFERENCE GROUPS BACK                                           22

Figure 1 Integrated Framework of Spiral of Silence

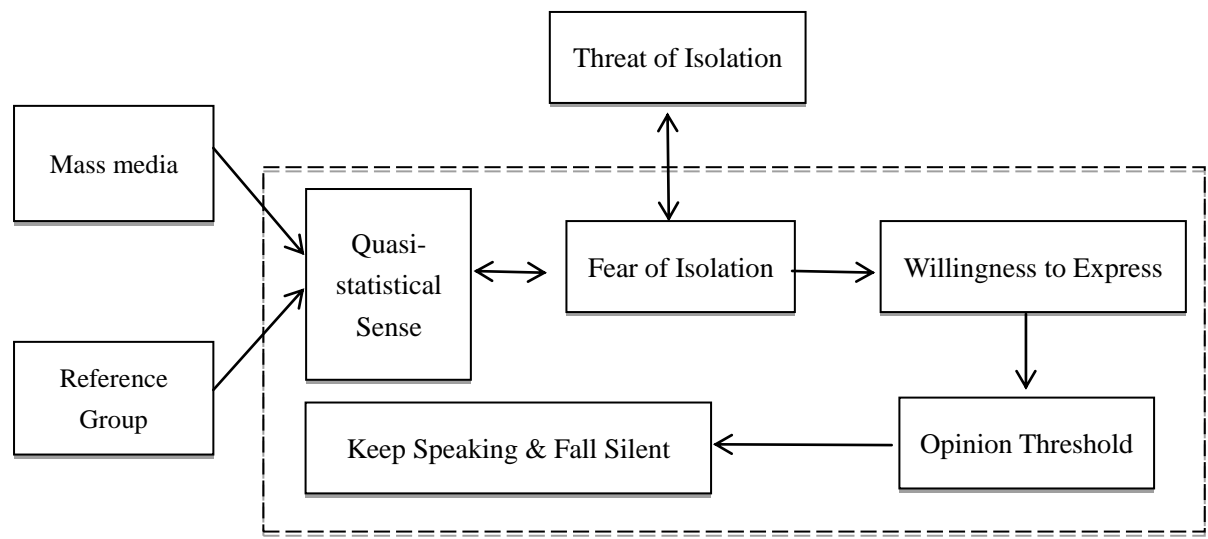

*Figure 2* Convergence of adjacent agents

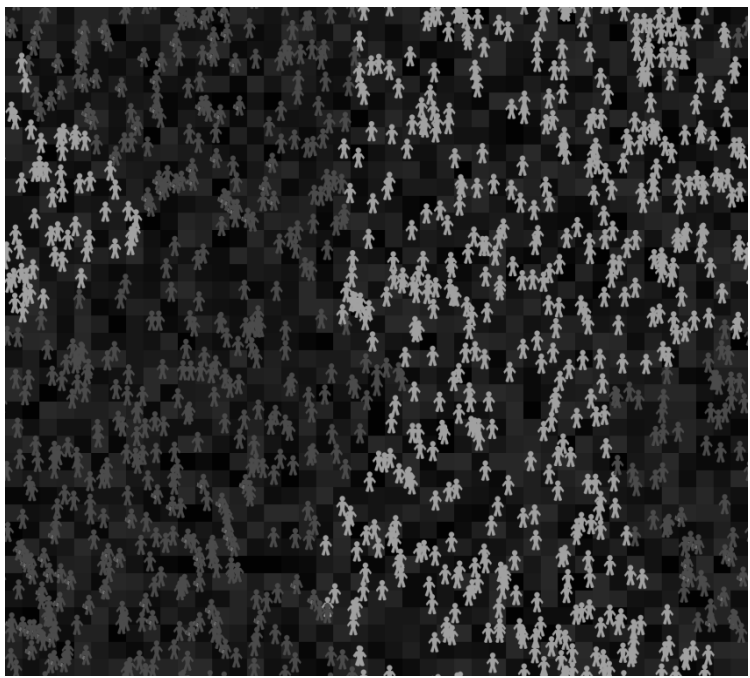